%% file: ANA-EXOT-2020-28-PAPER.tex
\documentclass[cernpreprint, atlasdraft=true, texlive=2020, USenglish,coverpage=false, texmf]{atlasdoc}

\usepackage[subfig]{atlaspackage}
\usepackage{tabularx}
\usepackage{multirow}
\usepackage{makecell}

\usepackage{atlasbiblatex}
 
\usepackage{atlasphysics}
 
\graphicspath{{logos/}{figures/}}
 
\usepackage{ANA-EXOT-2020-28-PAPER-defs}

\AtlasTitle{Search for Lepton-Flavor Violation in $Z$-Boson Decays with $\tau$
Leptons with the ATLAS Detector}
 
\AtlasAbstract{
A search for lepton-flavor-violating $Z\to e\tau$ and $Z\to\mu\tau$ decays with $pp$
collision data recorded by the ATLAS detector at the LHC is presented.
This analysis uses $139\,\mbox{fb\(^{-1}\)}$ of Run~2
$pp$ collisions at $\sqrt{s}=13$~\ensuremath{\text{Te\kern -0.1em V}}\ and is combined with the results of a similar ATLAS
search in the final state in which the $\tau$ lepton decays hadronically, using the same data set as well as Run~1 data.
The addition of leptonically decaying $\tau$ leptons
significantly improves the sensitivity reach for $Z\to\ell\tau$ decays.
The $Z\to\ell\tau$ branching fractions are constrained in this analysis
to $\mathcal{B}(Z\to e\tau)<7.0\times10^{-6}$
and $\mathcal{B}(Z\to \mu\tau)<7.2\times10^{-6}$ at 95\% confidence level.
The combination with the previously published analyses sets the strongest constraints
to date:
$\mathcal{B}(Z\to e\tau)<5.0\times10^{-6}$
and $\mathcal{B}(Z\to \mu\tau)<6.5\times10^{-6}$ at 95\% confidence level.
}
 
\author{ATLAS Collaboration}
 
\AtlasRefCode{EXOT-2020-28} 
\PreprintIdNumber{CERN-EP-2021-067}

\AtlasJournalRef{Physical Review Letter 127, 271801 (2021)}
\AtlasDOI{10.1103/PhysRevLett.127.271801}
\hypersetup{pdftitle={ATLAS document},pdfauthor={The ATLAS Collaboration}}
 
\begin{document}
 
\maketitle

\clearpage
 
Three lepton families (flavors) exist in the standard model of particle physics (SM)~\cite{SM1,SM2,SM3,SM4},
and the number of leptons of each family is conserved in their interactions.
Nevertheless, this conservation is not postulated by any fundamental principle
of the theory,
and neutrino oscillations~\cite{NO1,NO2} indicate that processes
violating this conservation do occur in nature.
According to current knowledge,  lepton-flavor-violating (LFV) processes in
charged-lepton interactions can occur via neutrino mixing but are too rare to
be detected by current experiments~\cite{ZLFV-predictions}.
An observation of these would be an unambiguous sign of physics beyond the SM.
LFV processes occur,  for example, in models predicting the existence of
heavy neutrinos~\cite{HN},  which may also explain the observed tiny masses
and large mixing of the SM neutrinos.  In such models, up to one in $10^{5}$
$Z$ bosons would undergo an LFV decay involving $\tau$ leptons.
In an earlier analysis, the ATLAS experiment at the LHC set the strongest constraints on the branching fractions
($\mathcal{B}$) of the LFV decays of the $Z$ boson involving a $\tau$ lepton by searching for such
decays in which the $\tau$ lepton decays hadronically~\cite{LFVpreprint}.  This result was achieved by analyzing proton-proton ($pp$) collision
data corresponding to an integrated luminosity of $139\,\ifb$ at a center-of-mass energy
$\sqrt{s}=13$~\TeV\ and $20.3\,\ifb$ at $\sqrt{s}=8$~\TeV.
In that search,  ATLAS measured the branching fractions to be $\mathcal{B}(Z\to e\tau) < 8.1\times10^{-6}$
and $\mathcal{B}(Z\to\mu\tau) < 9.5\times10^{-6}$ at 95\% confidence level (C.L.), superseding
former limits set by the LEP experiments of $\BR(Z\to e\tau) < 9.8\times10^{-6}$~\cite{LEP-Opal}
and $\BR(Z\to\mu\tau) < 1.2\times10^{-5}$~\cite{LEP-Delphi} at 95\% C.L.

This Letter presents a complementary search for $Z\to\ell\tau$ decays ($\ell=$
light charged lepton, i.e.\,$e$ or $\mu$) in which the $\tau$ leptons decay
into electrons or muons ($\ell\tau_{\mathrm{\ell^\prime}}$ channel) using $139\,\ifb$ of $pp$ collision data at $\sqrt{s}=13$~\TeV\
collected by the ATLAS experiment~\cite{PERF-2007-01,ATLAS-CONF-2019-021,LUCID2}.
The search is performed here for the first
time at the LHC and is combined with the similar ATLAS search using hadronic $\tau$-lepton
decays ($\ell\tau_{\mathrm{had}}$ channel)~\cite{LFVpreprint}.
The two searches follow similar analysis strategies.
Neural network classifiers are used for optimal discrimination of signal from backgrounds
and their distributions are employed in a binned maximum-likelihood fit to achieve better sensitivity.

ATLAS is a multipurpose particle
detector with a forward-backward symmetric cylindrical geometry and a near
$4\pi$ coverage in solid angle~\cite{PERF-2007-01,ATLAS-TDR-2010-19,PIX-2018-001}.
It consists of an inner tracking detector surrounded by a superconducting
solenoid, electromagnetic and hadronic calorimeters, and a muon spectrometer
based on superconducting air-core toroidal magnets.
This search analyzes $pp$ collision events recorded by the ATLAS experiment
using single-electron or single-muon triggers~\cite{TRIG-2016-01, TRIG-2018-05, TRIG-2018-01}.
Prompt electrons and muons from the $Z$-boson decays and those from the $\tau$-lepton decays are reconstructed and selected in the same way.
Candidates for electrons~\cite{EGAM-2018-01},  muons~\cite{MuonFullRun2},  jets~\cite{Cacciari:2008gp,Fastjet,JETM-2018-05},
and visible decay products of hadronic $\tau$-lepton decays (\tauhadvis)~\cite{ATL-PHYS-PUB-2019-033,ATLAS-CONF-2017-029} are reconstructed from energy deposits in the calorimeters and charged-particle tracks measured in the inner detector and the muon spectrometer.
These candidates are selected with sets of requirements similar to those used in Ref.~\cite{LFVpreprint}.
Electron candidates are required to pass the \emph{medium} likelihood-based identification requirement~\cite{EGAM-2018-01} and have a transverse momentum $\pt > 15$~\GeV\ and a pseudorapidity  $|\eta| < 1.37$ or $1.52<|\eta| < 2.47$.
The latter selection vetoes electron candidates passing through the transition
region between the barrel and end-cap electromagnetic calorimeters.
Muon candidates are required to pass the \emph{medium} identification requirement~\cite{PERF-2015-10} and have a $\pt > 10$~\GeV\ and $|\eta| < 2.5$.
Both the electron and muon candidates must satisfy the \emph{tight} isolation
requirement~\cite{EGAM-2018-01,PERF-2015-10}
, which is intended to reject misidentified candidates produced from the
hadronization of quarks or gluons based on tracks and clusters reconstructed
collinear to the candidates.
Events with exactly one electron and one muon candidate are selected with the requirement that the lepton with higher transverse momentum has a $\pt > 27$~\GeV.
This selection lies above the threshold for constant efficiency of both single-lepton trigger selections.
Events with same-flavor lepton pairs are rejected, in order to reduce the background from $Z\to\ell\ell$ decays.
Events with a leading-\pt electron are used in the search for $Z\to e\tau$ decays ($e\tau_{\mu}$ channel),
while those with a leading-\pt muon are used in the search for $Z\to \mu\tau$ decays ($\mu\tau_{e}$ channel),
assuming the prompt lepton from the $Z$-boson decay is the leading one in \pt.
In the $\mu\tau_{e}$ channel, the ratio of the electron's \pT reconstructed in the inner tracking detector
to the transverse energy reconstructed in the electromagnetic calorimeter, $\pT^\mathrm{track}(e)/\ET^\mathrm{cluster}(e)$,
is required to be smaller than $1.1$ in order to reject $Z\to\mu\mu$ events.
Opposite-charge lepton-pair events are analyzed in the search for signal events,  while events with same-charge lepton pairs are used for estimates of background processes.  Quark- or gluon-initiated particle showers (jets) are reconstructed using the anti-$k_t$ algorithm~\cite{Cacciari:2008gp,Fastjet} with a radius parameter $R = 0.4$.
Jets fulfilling $\pT >20$~\GeV\ and $|\eta| < 2.5$ are identified as
containing $b$ hadrons if tagged by a dedicated multivariate algorithm~\cite{FTAG-2018-01}.
To ensure the samples of selected events do not overlap with those used in the $\ell\tau_{\mathrm{had}}$ channel,
events with a \tauhadvis candidate are vetoed.
The \tauhadvis candidates reconstructed
from jets with a $\pT > \SI{10}{\GeV}$ and with one or three associated tracks are selected in $|\eta| < 1.37$ or $1.52<|\eta| < 2.5$.
The \tauhadvis identification is performed by a recurrent neural network algorithm~\cite{ATL-PHYS-PUB-2019-033}.
A \tauhadvis candidate is required to have a $\pt>25$~\GeV\ and pass the \emph{tight} identification selection.
The missing transverse momentum ($\boldsymbol{\MET}$) is calculated as
the negative
$\boldsymbol{\pT}$ sum
of all fully reconstructed and calibrated physics objects~\cite{PERF-2016-07,ATLAS-CONF-2018-023}.
Additionally, the calculation includes inner detector tracks that originate from the vertex associated with the hard-scattering process but are not associated with any of the reconstructed objects.

The $Z\to\ell\tau\to\ell\ell^{'}+2\nu$ 
signal events are characterized by a final state
which has two light charged leptons with different flavor and opposite electric charge,
two neutrinos and an invariant mass of all these particles
compatible with the $Z$-boson mass.
In most cases, these two leptons are emitted
approximately back-to-back in the plane transverse to the proton beam direction.
Since the $\tau$ lepton is typically boosted due to
the large difference between its mass and the mass of its parent $Z$ boson,
the two neutrinos from its decay are usually almost collinear with the charged
lepton from the $\tau$-lepton decay.
The dominant background contribution is from the lepton-flavor-conserving
$Z\to\tau\tau\to\ell\ell^{'}+4\nu$ decays, where the two $\tau$-leptons decay
leptonically. Subleading background contributions from other SM processes with
final states with two prompt leptons include the decays of a top-antitop-quark
pair ($t\bar{t}$), two gauge bosons (diboson), or a Higgs boson. Finally,
small background contributions come from $Z\to\ell\ell$ decays, where one of
the  light charged leptons is misidentified with the wrong flavor, and events
with ``fake leptons''. 
The latter type of background events 
includes mostly $W(\to\ell\nu)\text{+jets}$ events with 
leptons from heavy-flavor quark decays or with light-quark-initiated jets
that are misidentified as electrons or muons.
The signal and background events are separated by using a set of selection
criteria that define a signal-enhanced sample, referred to as the signal region (SR).
The selection criteria are listed in Table~\ref{tab:region_defs}.
Three neural network (NN) binary classifiers similar to those used in
Ref.~\cite{LFVpreprint} are trained on simulated events to distinguish signal
events from \Ztt, top-quark pair, and diboson background events individually.
The input to these NNs is a mixture of low- and high-level
kinematic variables,  following the same strategy as in the $\ell\tau_{\mathrm{had}}$ channel~\cite{LFVpreprint}.
The low-level variables are the momentum components of the reconstructed
electron and muon candidates, and the $\boldsymbol{\MET}$.
The high-level variables are kinematic properties of the
$e$-$\mu$-\MET system,
such as the collinear mass $\mcoll(e, \mu)$, defined as the invariant mass of the $e$-$\mu$-$2\nu$
system, where the two neutrinos are assumed to have a vectorial momentum
sum that is equal in \pT and the azimuthal angle $\phi$ around the beam axis to the measured $\boldsymbol{\MET}$ and equal in
$\eta$ to the subleading-\pt lepton momentum.
The outputs of the individual NNs ($\text{NN}_{i}$ with values between zero and one) are combined into a final
discriminant as shown in Eq.~(\ref{eq:NNcombFormula}), hereafter referred to as the ``combined NN output'':
\begin{equation}
\text{combined NN output} = 1- \sqrt{\frac13 \sum_{i=1}^3\left(1-\text{NN}_{i}\right)^2} \,.
\label{eq:NNcombFormula}
\end{equation}
Events classified by the NN trained for $\Ztt$ as backgroundlike are excluded
from the SR and used in a control region to better determine the \Ztt
background in the maximum-likelihood fit (see Table~\ref{tab:region_defs}).
The signal acceptance in the SR is
19.5\% for the $e\tau_{\mu}$ channel and 11.2\% for the $\mu\tau_e$ channel, as determined from simulated signal samples.
The lower acceptance in the $\mu\tau_e$ channel is due to the higher \pT-threshold on the subleading-\pt lepton and the additional selection on $\pT^\mathrm{track}/\ET^\mathrm{cluster}$.
 
\begin{table}[t]
\caption{Selection criteria for events in the signal region. The invariant
transverse mass of $\ell$ and $\boldsymbol{\MET}$\ is defined as
$m_{\mathrm{T}}(\ell,\boldsymbol{\MET})=\sqrt{2\,p_{\mathrm{T}}(\ell)\,\MET
\left[1-\cos\left(\phi_{\ell}-\phi_{\MET}\right)\right]}$. }
\label{tab:region_defs}
\centering
\begin{tabular}{p{0.59\textwidth}|p{0.36\textwidth}}
\toprule
\textbf{Selection criterion} & \textbf{Purpose} \\
\toprule
Exactly two isolated light leptons $(\ell_0,\ell_1)$ with opposite electric
charge and different flavor ($e$ or $\mu$); $\pT(\ell_{0})>\pT(\ell_{1})$ &
\multirow{2}{0.36\textwidth}{Select events consistent with signal decays. }\\
\midrule
No \tauhadvis candidate & Complementarity to the $\ell\tau_{\mathrm{had}}$
channel. \\
\midrule
Transverse mass $m_{\mathrm{T}}(\ell_{1},\boldsymbol{\MET}) < \SI{35}{\GeV}$ & \multirow{3}{0.36\textwidth}{Reject top-quark and diboson events.} \\
$|\Delta \phi(\ell_{0}, \boldsymbol{\MET})|>\SI{1}{\radian}$ & \\
No $b$-tagged jets (using the 77\% efficiency working point~\cite{FTAG-2018-01}) & \\
\midrule
Invariant mass of the $\ell_{0}$--$\ell_{1}$ pair $ m(\ell_{0},\ell_{1}) > \SI{40}{\GeV}$ &  Reject events incompatible with $Z$-boson decays. \\
\midrule
Neural network (optimized for signal vs.\ \Ztt) output $>$ 0.2 & Complementarity to the CRZ$\tau\tau$ region.\\
\midrule
In $\mu\tau_e$ channel: $\pT^\mathrm{track}(e)/\ET^\mathrm{cluster}(e)<1.1$ & Reject $Z\to\mu\mu$ events.\\
\bottomrule
\end{tabular}
\end{table}

Predictions for signal and background contributions are based partly on Monte Carlo (MC) simulations and
partly on estimates from data.  Signal and background processes were simulated as in Ref.~\cite{LFVpreprint}.
The signal events were simulated using \Pythia~8~\cite{Sjostrand:2014zea} with matrix elements calculated
at leading order (LO) in the strong coupling constant.
Nominal signal samples were generated with
a parity-conserving $Z\ell\tau$ vertex and unpolarized $\tau$ leptons.
Scenarios where the decays
are maximally parity violating were considered by reweighting the simulated events using \TauSpinner~\cite{TauSpinner},  as discussed in Ref.~\cite{LFVpreprint}.
The \Ztt background events were simulated with the \Sherpa~2.2.1~\cite{Bothmann:2019yzt} generator
using the NNPDF~3.0 NNLO PDF set~\cite{Ball:2014uwa}
and next-to-leading-order (NLO) matrix elements for up to two partons,
and LO matrix elements for up to four partons, calculated with the \Comix~\cite{Gleisberg:2008fv}
and \OpenLoops~\cite{Buccioni:2019sur,Cascioli:2011va,Denner:2016kdg} libraries.
Background \Zll events were simulated using
the \PowhegBox~\cite{Alioli:2010xd} generator with NLO matrix elements.
All MC samples include a detailed simulation of the ATLAS detector
with $\GEANT$~\cite{SOFT-2010-01,Agostinelli:2002hh}.
As in Ref.~\cite{LFVpreprint}, the simulation of $Z$-boson production is improved through a correction derived
from measurements in data. The simulated \pT spectra of the $Z$ boson are
reweighted to match the unfolded distribution measured by ATLAS in Ref.~\cite{STDM-2018-14}.
The predicted overall yields of signal and \Ztt events
are determined by a binned maximum-likelihood fit to the combined data
in the SR and in a control region enhanced in $Z \to \tau\tau$ events (\CRZtt).
This
eliminates the theoretical uncertainties in the total $Z$-boson production cross section ($\sigma_{Z}$),
as well as the experimental uncertainties related to the acceptance of the common
$\ell\ell^{'}$ final state.
The selection criteria for
events in the \CRZtt are the same as those for events in the SR, except that events
are required to be classified as $\Ztt$-like, i.e.~with an output smaller than 0.2 for the \Ztt NN and greater than 0.2 for both the top-quark and diboson NNs.
In the $\mu\tau_e$ channel,  a  small contribution to the total background
originates from $Z\to\mu\mu$ events
in which one muon is misreconstructed as an electron.
Such electron candidates may originate from muons that fail the muon selection
requirements and whose tracks are associated with a calorimeter energy cluster
and reconstructed as electrons. They may also originate from muons
undergoing bremsstrahlung.
Such events are modeled with simulation and their predicted yield is based on
the measured $\sigma_{Z}$~\cite{STDM-2015-03}.
The modeling is validated in a dedicated region which has the same selection
as the $\mu\tau_e$ SR except for the inverse selection on
$\pT^\mathrm{track}(e)/\ET^\mathrm{cluster}(e)$.
Based on the observed level of agreement between data and simulation,  a systematic uncertainty of 15\% is assigned to the predicted yield of $Z\to\mu\mu$ events in the SR, with no further correction.
 
Events with fake leptons yield a small but still significant background
contribution.  In most cases, the fake lepton is the subleading one.
These events are estimated from data using a ``fake-factor method''  similar to the one used in Ref.~\cite{LFVpreprint}.
The fake factor is defined as the ratio
$N_{\mathrm{fake}}^{\mathrm{pass-iso}}/N_{\mathrm{fake}}^{\mathrm{fail-iso}}$,
where ``fake'' indicates events with at least one fake lepton and
``pass-iso'' or ``fail-iso'' indicate whether the subleading lepton passes or fails the isolation requirement.
The fake factor is measured in events with pairs of same-sign leptons (SS).  These events are enhanced in $W(\to\ell\nu)$+jets, which is the dominant source of events with fake leptons in the SR.  Events in the SS region pass the same event selections as those in the SR except for a same-charge requirement.
The fake factors are measured as functions of the transverse momentum and pseudorapidity of the leptons,  separately for $e\tau_{\mu}$ and $\mu\tau_e$ events.
The kinematic properties of events with fake leptons in the SR or in the CRs are estimated by the distributions of events with the subleading lepton failing the isolation requirement, but otherwise satisfying all other selection criteria for that region, multiplied by the fake factor.
The total predicted yields of the events with fake leptons in the SR and CRs are instead
determined by a combined maximum-likelihood fit to data, separately for $e\tau_{\mu}$ and $\mu\tau_e$ events.
The remaining background processes are estimated using simulations.
These backgrounds include events from the production and decay of top quarks~\cite{Alioli:2010xd,Sjostrand:2014zea},
pairs of gauge bosons~\cite{Bothmann:2019yzt, Ball:2014uwa}\ and the Higgs boson~\mbox{\cite{Alioli:2010xd,Sjostrand:2014zea}}.  The yield of the events with
top quarks is determined in the maximum-likelihood fit to data via the
inclusion of a top-quark control region (CRTop).
The selection requirements for the CRTop are the same as for the SR except that at least one $b$-tagged jet is required.  
The expected event yields of the remaining processes are determined based on
their production cross section, the integrated luminosity, and the simulated
selection efficiency.

A statistical analysis of the selected events is performed to
assess the presence of signal events, following the same method used in Ref.~\cite{LFVpreprint}.
A simultaneous binned maximum-likelihood fit to the combined NN output
distribution in the SR,  the $\mcoll(e, \mu)$ distribution in the \CRZtt, and the event yield in CRTop is used to constrain uncertainties in the predictions and extract evidence of a possible signal.
The fit is performed independently for the $e\tau$ and $\mu\tau$ channels.
The fraction of $Z\to e\tau$ events selected in the $\mu\tau$ channel (and vice
versa) is negligible and is therefore neglected.
In order to improve the discrimination between signal and the events with fake
leptons,  the events in the SR are further split into two regions based on the
transverse momentum of the subleading-\pt lepton $\ell_{1}$.  The low-\pt SR contains events with a $\pt(\ell_{1})< 20\,(25)$~\GeV\ in the $e\tau_{\mu}$ ($\mu\tau_e$) channel, while the high-\pt SR contains the events above these thresholds.
Both SRs in the
$e\tau_{\mu}$ channel have comparable sensitivity, while the
low-\pT SR in the $\mu\tau_e$ channel is more sensitive than the high-\pT SR.
Both SRs are fitted simultaneously.
There are four unconstrained parameters in the fits: the parameter of interest determines the LFV branching fraction $\BR(Z\to\ell\tau)$ by modifying an arbitrary prefit signal yield, $\mu_{Z}$ determines $\sigma_{Z}$ times the overall acceptance
and reconstruction efficiency of the $\ell\ell^{'}$ final state in $Z\to\tau\tau$ and signal events, $\mu_{\mathrm{top}}$ determines the yield of the top-quark events, and $\mu_{\mathrm{fakes}}$ determines the yield of the events with fake leptons.
Constrained parameters are also introduced to account for
systematic uncertainties in the signal and background predictions, as in Ref.~\cite{LFVpreprint}.
These include uncertainties in simulated events in the modeling of trigger, reconstruction, identification
and isolation efficiencies,  as well as energy calibrations and resolutions of reconstructed objects.
No systematic uncertainties are assigned to the overall yields of events with $Z$-boson decays,
fake leptons, or top quarks as these yields are determined from data.
Uncertainties related to events with fake leptons include statistical uncertainties due to the size of
the data sample used to measure the fake factors as well as to model their distributions in the SRs and CRs.
Systematic uncertainties assigned to events with fake leptons account for: shape differences in the
modeling of the combined NN output in the SS events;
differences in the composition of the events with fake leptons between SS events and the events in the SRs;
and uncertainties affecting the number of events with prompt leptons failing the isolation requirements as estimated by simulation. 
The dominant uncertainties of the search are statistical in nature.  Among the systematic uncertainties, the dominant ones are those in the jet calibration which enter through the calculation of the $\boldsymbol{\MET}$~\cite{JETM-2018-05}.
A summary of the uncertainties and their impact on the LFV branching fraction is given in Table~\ref{tab:syst_breakdown}.

\begin{table}[t]
\centering
\caption{Summary of the contributions to the uncertainty in the measured $\BR(Z\to\ell\tau_{\ell^{\prime}})$.
The uncertainties related to light charged leptons include those in the trigger, reconstruction, identification and isolation
efficiencies, as well as energy calibrations.
The uncertainties related to jets and $\boldsymbol{\MET}$ include those in the energy
calibration and resolution.
The uncertainty in the $Z\to\mu\mu$ yield is only applicable in the
$\mu\tau$ channel.
The total systematic uncertainty can differ from the sum in quadrature of the different contributions due to correlations among uncertainties as a result of the likelihood fit to data.
}
\label{tab:syst_breakdown}
\begin{tabularx}{0.475\textwidth}{lYY}
\toprule
\multicolumn{3}{r}{\textbf{Uncertainty in $\BR(\Zlt)$ [$\times10^{-6}$]}} \\
\textbf{Source of uncertainty} & \textbf{$e\tau$} & \textbf{$\mu\tau$} \\
\midrule
Statistical                                         & $\pm$3.5 & $\pm$3.9 \\
\hspace{3ex} Fake leptons (statistical)                    & $\pm$0.1 & $\pm$0.1 \\
Systematic                                          & $\pm$2.7 & $\pm$3.4 \\
\hspace{3ex} Light charged leptons                          & $\pm$0.4 & $\pm$0.4 \\
\hspace{3ex} $\boldsymbol{\MET}$                       & $\pm$0.4 & $\pm$0.8 \\
\hspace{3ex} Jets                       & $\pm$1.9 & $\pm$2.2 \\
\hspace{3ex} Flavor tagging            & $\pm$0.5 & $\pm$0.9 \\
\hspace{3ex} $Z$-boson modeling                    & $<$0.1 & $\pm$0.1 \\
\hspace{3ex} $Z\to\mu\mu$ yield                   &  & $\pm$0.8 \\
\hspace{3ex} Other  backgrounds                   & $\pm$0.1 & $\pm$0.6 \\
\hspace{3ex} Fake leptons (systematic)         & $\pm$0.4 & $\pm$0.9 \\
\midrule
Total                                               & $\pm$4.4 & $\pm$5.2 \\
\bottomrule
\end{tabularx}
\end{table}
 
\begin{figure}[htbp]
\centering
\subfloat[]{
\includegraphics[width=0.4\textwidth]{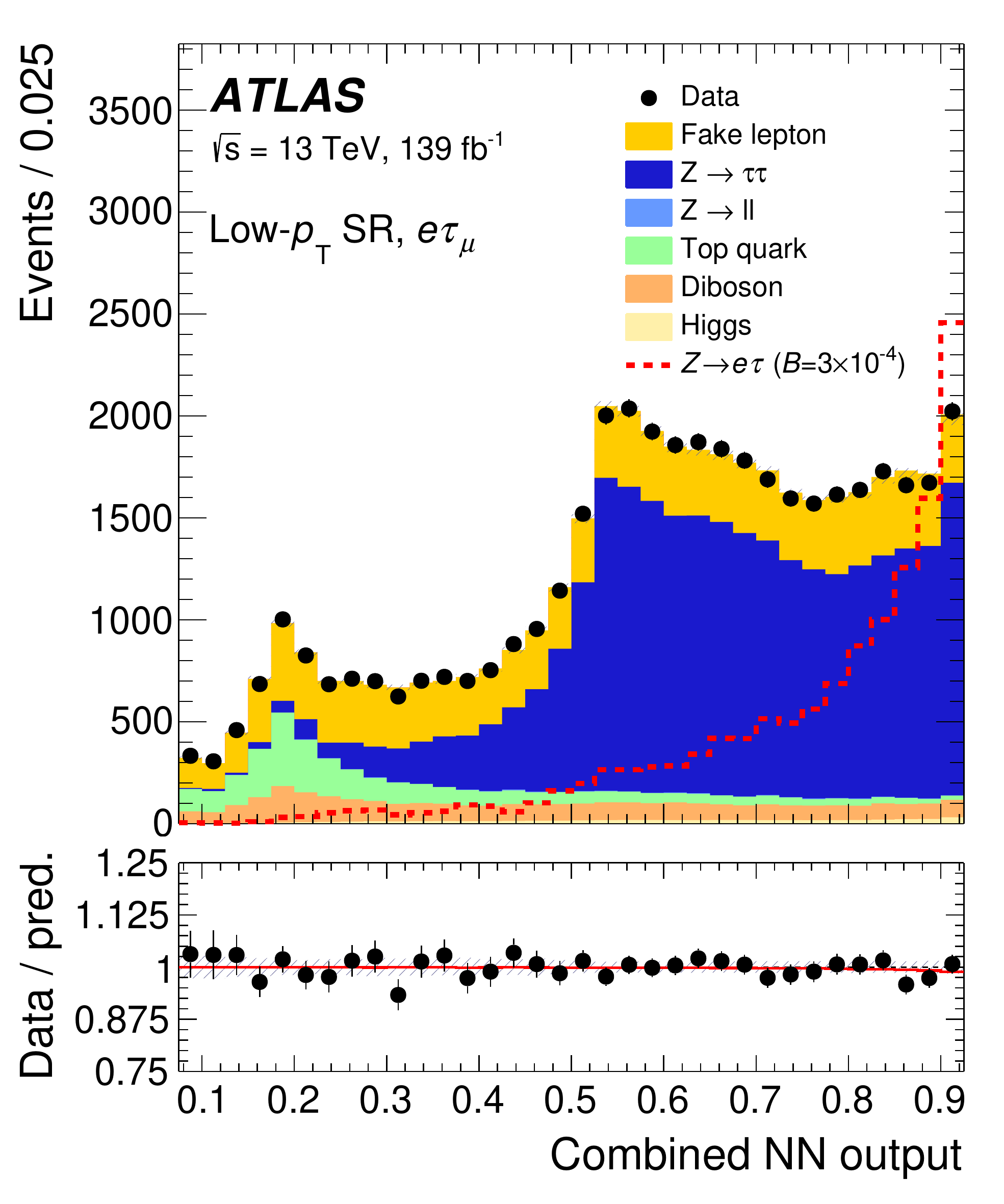}
}
\subfloat[]{
\includegraphics[width=0.4\textwidth]{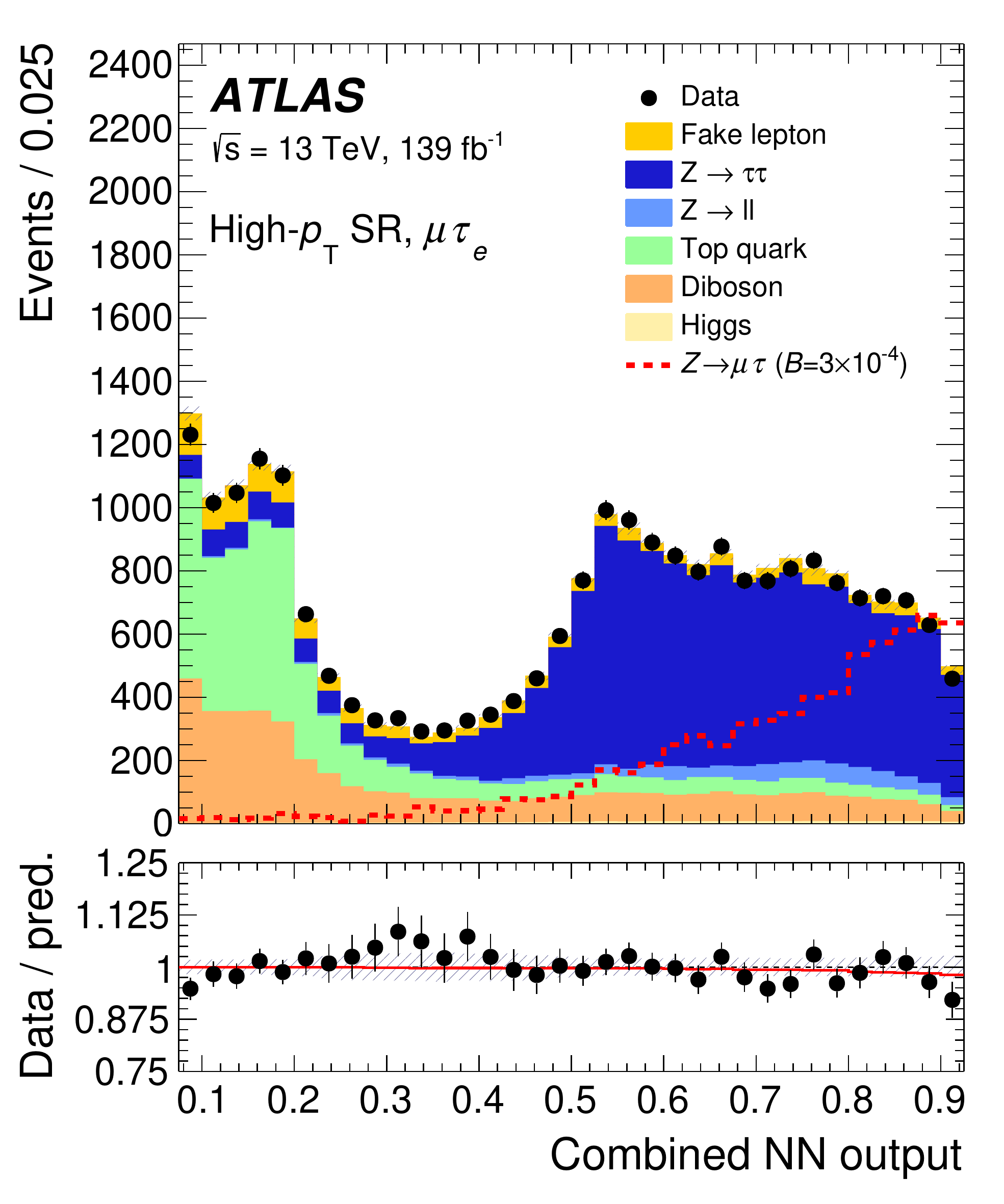}
}\\
\subfloat[]{
\includegraphics[width=0.4\textwidth]{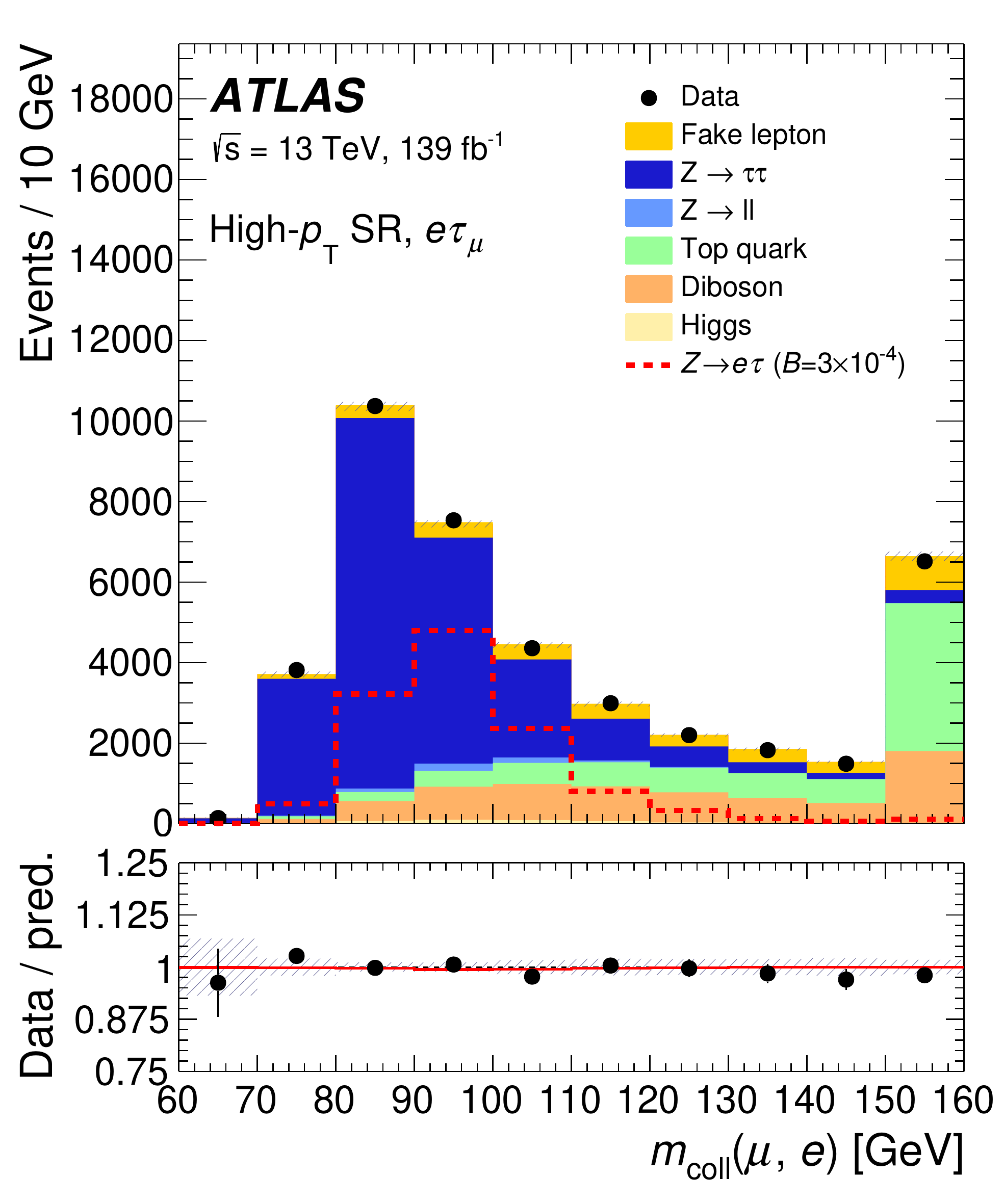}
}
\subfloat[]{
\includegraphics[width=0.4\textwidth]{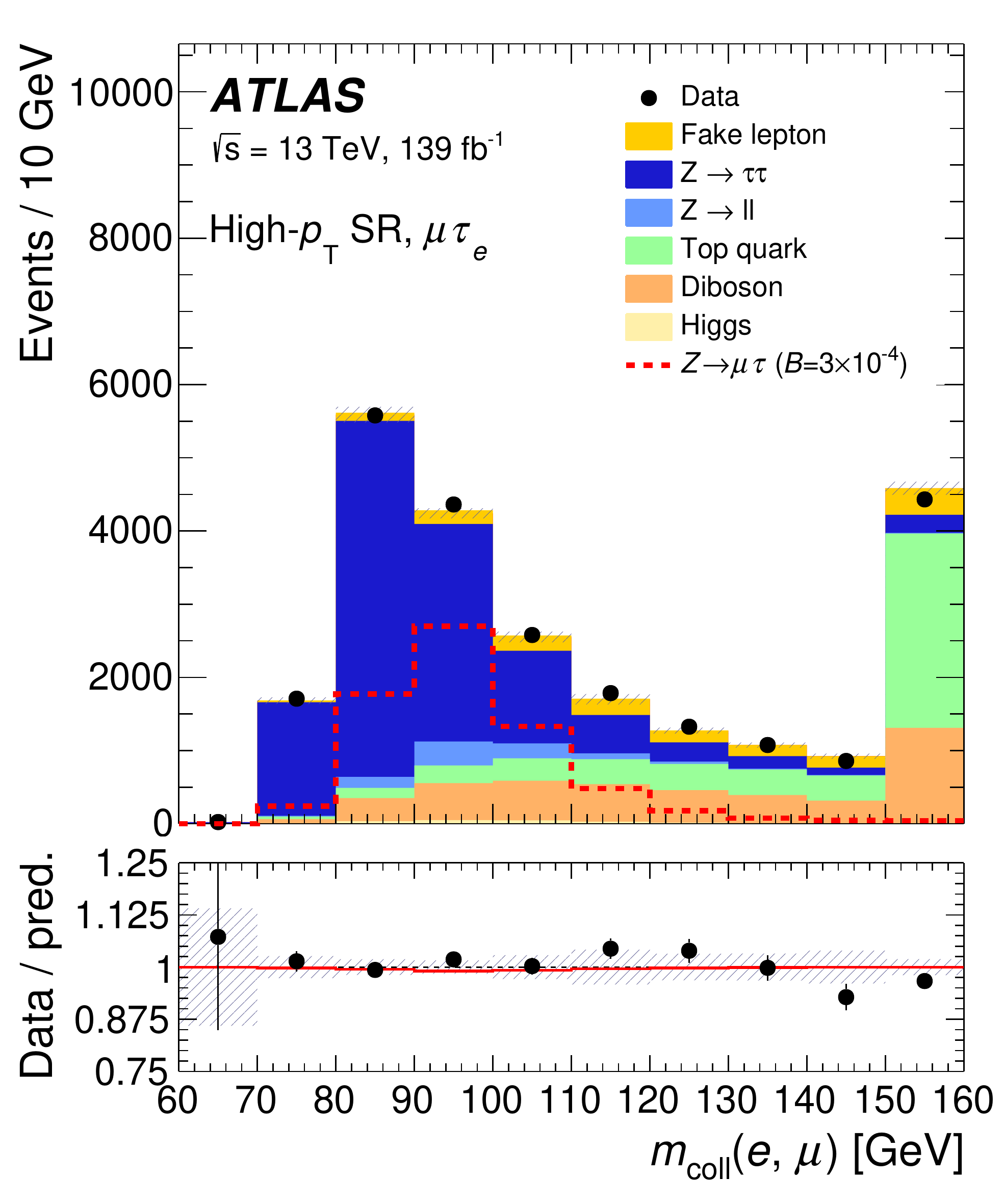}
}
\caption{
Observed and best-fit predicted distributions in the SRs.
Distributions of the combined NN output are shown in (a) for the
low-\pT SR of the $e\tau_\mu$ channel, and in (b) for the high-\pT SR of
the $\mu\tau_e$ channel.
Distributions of the collinear mass in the high-\pT SR are shown in
(c) and (d) for the $e\tau_\mu$ and $\mu\tau_e$ channels,
respectively.
The expected signal, normalized to an arbitrary
$\BR(Z\to\ell\tau)=3{\times}10^{-4}$ for visualization purposes, is shown as
a dashed histogram in each plot.
In the panel below each plot, the ratios of the observed yield (dots) and
the best-fit background-plus-signal yield (solid line) to the best-fit
background yield are shown.
The hatched uncertainty bands represent one standard deviation of the combined
statistical and systematic uncertainties.
The first and last bins in each plot include underflow and overflow events,  respectively.
}
\label{fig:postfit_dists}
\end{figure}
 
The observed and best-fit predicted distributions of the combined NN output in
the SRs with the highest sensitivity
as well as distributions of the collinear mass in the high-\pt SRs
are shown in Fig.~\ref{fig:postfit_dists}.
The best-fit yield of \Zlt signal corresponds to the branching fractions
$\BR(\Zet)=[-2.6\pm3.5\,(\text{stat}) \pm 2.7\,(\text{syst})]\times10^{-6}$ and
$\BR(\Zmt)=[-4.4\pm3.9\,(\text{stat}) \pm 3.4\,(\text{syst})]\times10^{-6}$.
The best-fit yields of \Ztt, top quarks, and events with fake leptons are close to the prefit predicted values
and are determined with a relative precision of 2\%-4\%,  except the events with fake leptons in the $\mu\tau_e$
channel, which have an uncertainty of 30\%.
As no significant excess of data over the predicted background is observed, a
combined fit of the $\ell\tau_{\mathrm{\ell^\prime}}$ and
$\ell\tau_{\mathrm{had}}$ channels is used to set upper limits on $\BR(\Zlt)$.
The analysis of the $\ell\tau_{\mathrm{had}}$ channel with Run~2 data~\cite{LFVpreprint} uses a similar scheme of regions and unconstrained parameters.
In the statistical combination,  the parameters of interest are correlated
among the different SRs and CRs.
The other unconstrained parameters are
uncorrelated as these account either for backgrounds specific to each channel or for
different acceptances of the $\ell\tau_{\ell^{'}}$ or $\ell\tau_{\mathrm{had}}$ final states.
Common systematic uncertainties are correlated,  besides those related to the jet energy calibrations, which are
uncorrelated.  This conservative correlation scheme was chosen because of different best-fit values
for the parameters associated with these uncertainties in the two channels.
However,  the fit with correlated jet energy calibration uncertainties
yields compatible combined upper limits.  The analysis of the $\ell\tau_{\mathrm{had}}$ channel with Run~1 data is combined
using the same correlation scheme as in Ref.~\cite{LFVpreprint}.
The combined best-fit amount of \Zlt signal corresponds to the branching fractions
$\BR(\Zet)=[-1.4\pm 2.5\,(\text{stat}) \pm 1.8\,(\text{syst})]\times10^{-6}$ and
$\BR(\Zmt)=[1.7\pm 2.2\,(\text{stat}) \pm 1.6\,(\text{syst})]\times10^{-6}$.
 
\begin{table}[hbtp]
\centering
\caption{Observed and expected (median) upper limits on the signal branching fraction at 95\% C.L., in different $\tau$-polarization scenarios.}
\label{tab:ULs_unblind_tauSpin}
\begin{tabularx}{\textwidth}{lYY}
\toprule
\multicolumn{3}{r}{\textbf{Observed (expected) upper limit on $\BR(Z\to\ell\tau)$ [$\times10^{-6}$]}} \\
\textbf{Final state, polarization assumption} & \textbf{$e\tau$} & \textbf{$\mu\tau$} \\
\midrule
$\ell\tau_{\mathrm{had}}$ Run~1 + Run~2,  unpolarized $\tau$  \cite{LFVpreprint} & 8.1 (8.1) & 9.5 (6.1) \\
$\ell\tau_{\mathrm{had}}$ Run~2,  left-handed $\tau$  \cite{LFVpreprint} & 8.2 (8.6) & 9.5 (6.7) \\
$\ell\tau_{\mathrm{had}}$ Run~2,  right-handed $\tau$  \cite{LFVpreprint} & 7.8 (7.6) & 10 (5.8) \\
\midrule
$\ell\tau_{\mathrm{\ell^\prime}}$ Run~2, unpolarized $\tau$  & 7.0 (8.9) & 7.2 (10) \\
$\ell\tau_{\mathrm{\ell^\prime}}$ Run~2, left-handed $\tau$  & 5.9 (7.5) & 5.7 (8.5) \\
$\ell\tau_{\mathrm{\ell^\prime}}$ Run~2,  right-handed $\tau$ & 8.4 (11)  & 9.8 (13) \\
\midrule
Combined $\ell\tau$ Run~1 + Run~2,  unpolarized $\tau$  & 5.0 (6.0) & 6.5 (5.3) \\
Combined $\ell\tau$ Run~2, left-handed $\tau$  & 4.5 (5.7) & 5.6 (5.3) \\
Combined $\ell\tau$ Run~2, right-handed $\tau$ & 5.4 (6.2) & 7.7 (5.3) \\
\bottomrule
\end{tabularx}
\end{table}
\FloatBarrier
Since no significant deviation from the SM background hypothesis is observed,  exclusion limits are set using the \CLs method~\cite{CLs}.
The upper limits are shown in Table~\ref{tab:ULs_unblind_tauSpin} for LFV decays
with different assumptions about the $\tau$-polarization state.
The polarization of the $\tau$ lepton affects the energy of its visible decay
products and thus the acceptance for signal events.
In the scenario where the $\tau$ leptons are unpolarized, the observed upper limits at 95\% C.L. on
$\BR(\Zet)$ and $\BR(\Zmt)$ are $5.0\times 10^{-6}$ and $6.5\times 10^{-6}$, respectively.
 
In conclusion, this Letter reports the first analysis of the $\ell\tau_{\ell^\prime}$ channel
in the search for $Z\to\ell\tau$ decays at the LHC.  This channel
yields a sensitivity similar to the $\ell\tau_{\mathrm{had}}$ channel.
With the combined results of the two channels,  the ATLAS experiment sets the most stringent constraints on
LFV $Z$-boson decays involving $\tau$ leptons to date.
The precision of these results is mainly limited by statistical uncertainties.
 
\FloatBarrier
 
\section*{Acknowledgments}
 

We thank CERN for the very successful operation of the LHC, as well as the
support staff from our institutions without whom ATLAS could not be
operated efficiently.
 
We acknowledge the support of
ANPCyT, Argentina;
YerPhI, Armenia;
ARC, Australia;
BMWFW and FWF, Austria;
ANAS, Azerbaijan;
SSTC, Belarus;
CNPq and FAPESP, Brazil;
NSERC, NRC and CFI, Canada;
CERN;
ANID, Chile;
CAS, MOST and NSFC, China;
Minciencias, Colombia;
MEYS CR, Czech Republic;
DNRF and DNSRC, Denmark;
IN2P3-CNRS and CEA-DRF/IRFU, France;
SRNSFG, Georgia;
BMBF, HGF and MPG, Germany;
GSRI, Greece;
RGC and Hong Kong SAR, China;
ISF and Benoziyo Center, Israel;
INFN, Italy;
MEXT and JSPS, Japan;
CNRST, Morocco;
NWO, Netherlands;
RCN, Norway;
MEiN, Poland;
FCT, Portugal;
MNE/IFA, Romania;
JINR;
MES of Russia and NRC KI, Russian Federation;
MESTD, Serbia;
MSSR, Slovakia;
ARRS and MIZ\v{S}, Slovenia;
DSI/NRF, South Africa;
MICINN, Spain;
SRC and Wallenberg Foundation, Sweden;
SERI, SNSF and Cantons of Bern and Geneva, Switzerland;
MOST, Taiwan;
TAEK, Turkey;
STFC, United Kingdom;
DOE and NSF, United States of America.
In addition, individual groups and members have received support from
BCKDF, CANARIE, Compute Canada and CRC, Canada;
COST, ERC, ERDF, Horizon 2020 and Marie Sk{\l}odowska-Curie Actions, European Union;
Investissements d'Avenir Labex, Investissements d'Avenir Idex and ANR, France;
DFG and AvH Foundation, Germany;
Herakleitos, Thales and Aristeia programmes co-financed by EU-ESF and the Greek NSRF, Greece;
BSF-NSF and GIF, Israel;
Norwegian Financial Mechanism 2014-2021, Norway;
NCN and NAWA, Poland;
La Caixa Banking Foundation, CERCA Programme Generalitat de Catalunya and PROMETEO and GenT Programmes Generalitat Valenciana, Spain;
G\"{o}ran Gustafssons Stiftelse, Sweden;
The Royal Society and Leverhulme Trust, United Kingdom.
 
The crucial computing support from all WLCG partners is acknowledged gratefully, in particular from CERN, the ATLAS Tier-1 facilities at TRIUMF (Canada), NDGF (Denmark, Norway, Sweden), CC-IN2P3 (France), KIT/GridKA (Germany), INFN-CNAF (Italy), NL-T1 (Netherlands), PIC (Spain), ASGC (Taiwan), RAL (UK) and BNL (USA), the Tier-2 facilities worldwide and large non-WLCG resource providers. Major contributors of computing resources are listed in Ref.~\cite{ATL-SOFT-PUB-2021-003}.
 

\printbibliography

\clearpage \input{atlas_authlist}

\end{document}

%% file: atlas_authlist.tex
 
\begin{flushleft}
\hypersetup{urlcolor=black}
{\Large The ATLAS Collaboration}

\bigskip

\AtlasOrcid[0000-0002-6665-4934]{G.~Aad}$^\textrm{\scriptsize 99}$,    
\AtlasOrcid[0000-0002-5888-2734]{B.~Abbott}$^\textrm{\scriptsize 125}$,    
\AtlasOrcid[0000-0002-7248-3203]{D.C.~Abbott}$^\textrm{\scriptsize 100}$,    
\AtlasOrcid[0000-0002-2788-3822]{A.~Abed~Abud}$^\textrm{\scriptsize 34}$,    
\AtlasOrcid[0000-0002-1002-1652]{K.~Abeling}$^\textrm{\scriptsize 51}$,    
\AtlasOrcid[0000-0002-2987-4006]{D.K.~Abhayasinghe}$^\textrm{\scriptsize 91}$,    
\AtlasOrcid[0000-0002-8496-9294]{S.H.~Abidi}$^\textrm{\scriptsize 27}$,    
\AtlasOrcid[0000-0001-5329-6640]{H.~Abramowicz}$^\textrm{\scriptsize 158}$,    
\AtlasOrcid[0000-0002-1599-2896]{H.~Abreu}$^\textrm{\scriptsize 157}$,    
\AtlasOrcid[0000-0003-0403-3697]{Y.~Abulaiti}$^\textrm{\scriptsize 5}$
% ...
\bigskip
\\

% ...
$^{5}$High Energy Physics Division, Argonne National Laboratory, Argonne IL; United States of America.\\
% ...
$^{27}$Physics Department, Brookhaven National Laboratory, Upton NY; United States of America.\\
% ...
$^{34}$CERN, Geneva; Switzerland.\\
% ...
$^{51}$II. Physikalisches Institut, Georg-August-Universit\"{a}t G\"ottingen, G\"ottingen; Germany.\\
% ...
$^{91}$Department of Physics, Royal Holloway University of London, Egham; United Kingdom.\\
% ...
$^{99}$CPPM, Aix-Marseille Universit\'e, CNRS/IN2P3, Marseille; France.\\
$^{100}$Department of Physics, University of Massachusetts, Amherst MA; United States of America.\\
% ...
$^{125}$Homer L. Dodge Department of Physics and Astronomy, University of Oklahoma, Norman OK; United States of America.\\
% ...
$^{157}$Department of Physics, Technion, Israel Institute of Technology, Haifa; Israel.\\
$^{158}$Raymond and Beverly Sackler School of Physics and Astronomy, Tel Aviv University, Tel Aviv; Israel.\\
% ...

\end{flushleft}


%% file: ANA-EXOT-2020-28-PAPER.bib
@article{SM1,
    author = "Glashow, S.L.",
    title = {Partial-symmetries of weak interactions},
    doi = "10.1016/0029-5582(61)90469-2",
    journal = "Nucl. Phys.",
    volume = "22",
    pages = "579--588",
    year = "1961"
}

@article{SM2,
  title = {A Model of Leptons},
  author = {Weinberg, Steven},
  journal = {Phys. Rev. Lett.},
  volume = {19},
  %issue = {21},
  pages = {1264--1266},
  numpages = {0},
  year = {1967},
  month = {Nov},
  publisher = {American Physical Society},
  doi = {10.1103/PhysRevLett.19.1264}
  %url = {https://link.aps.org/doi/10.1103/PhysRevLett.19.1264}
}

@INBOOK{SM3,
       author = {{Salam}, Abdus},
        title = "{Weak and Electromagnetic Interactions}",
    booktitle = {Selected Papers of Abdus Salam},
    publisher = {World Scientific},
      address = {Singapore},
         year = 1994,
        pages = {{244-254}},
	  doi = {10.1142/9789812795915_0034}
}

@article{SM4,
  title = "Regularization and renormalization of gauge fields",
  journal = "Nucl. Phys. B",
  volume = "44",
  number = "1",
  pages = "189 - 213",
  year = "1972",
  %issn = "0550-3213",
  doi = "https://doi.org/10.1016/0550-3213(72)90279-9",
  %url = "http://www.sciencedirect.com/science/article/pii/0550321372902799",
  author = "G. {'t Hooft} and M. Veltman",
}

@article{NO1,
  title = {Evidence for Oscillation of Atmospheric Neutrinos},
  %author = {Fukuda, Y. and others},
  author = {{Super-Kamiokande Collaboration}},
  journal = {Phys. Rev. Lett.},
  volume = {81},
  %issue = {8},
  pages = {1562--1567},
  numpages = {0},
  year = {1998},
  month = {Aug},
  publisher = {American Physical Society},
  doi = {10.1103/PhysRevLett.81.1562},
  %url = {https://link.aps.org/doi/10.1103/PhysRevLett.81.1562}
}

@article{NO2,
  title = {Direct Evidence for Neutrino Flavor Transformation from Neutral-Current Interactions in the Sudbury Neutrino Observatory},
  %author = {Ahmad, Q. R. and others},
  author = {{SNO Collaboration}},
  journal = {Phys. Rev. Lett.},
  volume = {89},
  %issue = {1},
  pages = {011301},
  numpages = {6},
  year = {2002},
  month = {Jun},
  publisher = {American Physical Society},
  doi = {10.1103/PhysRevLett.89.011301},
  %url = {https://link.aps.org/doi/10.1103/PhysRevLett.89.011301}
}

@article{HN,
  title = {Charged lepton flavor violation from massive neutrinos in $Z$ decays},
  author = {Illana, J. I. and Riemann, T.},
  journal = {Phys. Rev. D},
  volume = {63},
  %issue = {5},
  pages = {053004},
  numpages = {13},
  year = {2001},
  month = {Feb},
  publisher = {American Physical Society},
  doi = {10.1103/PhysRevD.63.053004},
  %url = {https://link.aps.org/doi/10.1103/PhysRevD.63.053004}
}

@inproceedings{ZLFV-predictions,
    author = "Illana, Jose I. and Jack, M. and Riemann, T.",
    title = "{Predictions for $Z\rightarrow\mu\tau$ and Related Reactions}",
    booktitle = "{2nd Workshop of the 2nd Joint ECFA / DESY Study on Physics and Detectors for a Linear Electron Positron Collider}",
    eprint = "hep-ph/0001273",
    archivePrefix = "arXiv",
    reportNumber = "DESY-99-165, UG-FT-112-00, LC-TH-2000-007",
    pages = "490--524",
    month = "12",
    year = "1999"
}

@article{LEP-Delphi,
      %author         = "Abreu, P. and others",
      author         = "{DELPHI Collaboration}",
      title          = "{Search for lepton flavour number violating $Z^0$ decays}",
      %collaboration  = "DELPHI",
      journal        = "Z. Phys. C",
      volume         = "73",
      year           = "1997",
      pages          = "243-251",
      doi            = "10.1007/s002880050313",
      reportNumber   = "CERN-PPE-96-129",
      SLACcitation   = "%%CITATION = ZEPYA,C73,243;%%"
}

@article{LEP-Opal,
      %author         = "Akers, R. and others",
      author         = "{OPAL Collaboration}",
      title          = "{A search for lepton flavour violating $Z^0$ decays}",
      collaboration  = "OPAL",
      journal        = "Z. Phys. C",
      volume         = "67",
      year           = "1995",
      pages          = "555-564",
      doi            = "10.1007/BF01553981",
      reportNumber   = "CERN-PPE-95-043, CERN-PPE-95-43",
      SLACcitation   = "%%CITATION = ZEPYA,C67,555;%%"
}

@article{STDM-2018-14,  % Not yet in ATLAS.bib
    author         = "{ATLAS Collaboration}",
    title = "{Measurement of the transverse momentum distribution of Drell-Yan lepton pairs in proton-proton collisions at $\sqrt{s}=13$ TeV with the ATLAS detector}",
    eprint = "1912.02844",
    archivePrefix = "arXiv",
    primaryClass = "hep-ex",
    reportNumber = "CERN-EP-2019-223",
    doi = "10.1140/epjc/s10052-020-8001-z",
    journal = "Eur. Phys. J. C",
    volume = "80",
    number = "7",
    pages = "616",
    year = "2020"
}

@article{JETM-2018-05,  % Not yet in ATLAS.bib
    author         = "{ATLAS Collaboration}",
    title = "{Jet energy scale and resolution measured in proton-proton collisions at $\sqrt{s}=13$ TeV with the ATLAS detector}",
    eprint = "2007.02645",
    archivePrefix = "arXiv",
    primaryClass = "hep-ex",
    reportNumber = "CERN-EP-2020-083",
    month = "7",
    journal = "Eur. Phys. J. C",
    volume = "81",
    pages = "689",
    doi = "10.1140/epjc/s10052-021-09402-3",
    year = "2021"
}

@article{TauSpinner,
      author         = "Przedzinski, T. and Richter-Was, E. and Was, Z.",
      title          = "{Documentation of $TauSpinner$ algorithms: program for
                        simulating spin effects in $\tau$-lepton production at
                        LHC}",
      journal        = "Eur. Phys. J. C",
      volume         = "79",
      year           = "2019",
      number         = "2",
      pages          = "91",
      doi            = "10.1140/epjc/s10052-018-6527-0",
      eprint         = "1802.05459",
      archivePrefix  = "arXiv",
      primaryClass   = "hep-ph",
      reportNumber   = "IFJPAN-IV-2018-3",
      SLACcitation   = "%%CITATION = ARXIV:1802.05459;%%"
}

@article{CLs,
  author={A L Read},
  title={Presentation of search results: the CL$_\text{s}$ technique},
  journal={J. Phys. G},
  volume={28},
  doi = {10.1088/0954-3899/28/10/313},
  %issue={10},
  pages={2693},
  year={2002},
}

@article{LFVpreprint,
    author         = "{ATLAS Collaboration}",
    title          = "{Search for charged-lepton-flavour violation in Z-boson decays with the ATLAS detector}",
    journal        = "Nature Phys.",
    volume         = "17",
    year           = "2021",
    pages          = "819-825",
    doi            = "10.1038/s41567-021-01225-z",
    reportNumber   = "CERN-EP-2020-163",
    %eprint         = "2010.02566",
    %archivePrefix  = "arXiv",
    %primaryClass   = "hep-ex",
}

@article{LUCID2,
  author={G. Avoni and others},
  title={The new LUCID-2 detector for luminosity measurement and monitoring in ATLAS},
  journal={JINST},
  volume={13},
  number={07},
  pages={P07017},
  doi="10.1088/1748-0221/13/07/P07017",
  year={2018},
}

@article{MuonFullRun2,
    author = "{ATLAS Collaboration}",
    collaboration = "ATLAS",
    title = "{Muon reconstruction and identification efficiency in ATLAS using the full Run 2 $pp$ collision data set at $\sqrt{s}=13$ TeV}",
    eprint = "2012.00578",
    archivePrefix = "arXiv",
    primaryClass = "hep-ex",
    reportNumber = "CERN Preprint ID: CERN-EP-2020-199",
    month = "12",
    journal = "Eur. Phys. J. C",
    volume = "81",
    pages = "578",
    doi = "10.1140/epjc/s10052-021-09233-2",
    year = "2021"
}


%% file: ATLAS-useful.bib
@Article{Cacciari:2008gp,
     author    = "Cacciari, Matteo and Salam, Gavin P. and Soyez, Gregory",
     title     = "{The anti-\(k_{t}\) jet clustering algorithm}",
     journal   = "JHEP",
     volume    = "04",
     year      = "2008",
     pages     = "063",
     eprint    = "0802.1189",
     archivePrefix = "arXiv",
     primaryClass  =  "hep-ph",
     doi       = "10.1088/1126-6708/2008/04/063",
     SLACcitation  = "%%CITATION = 0802.1189;%%"
}

@Article{Fastjet,
      author         = "Cacciari, Matteo and Salam, Gavin P. and Soyez, Gregory",
      title          = "{FastJet user manual}",
      journal        = "Eur. Phys. J. C",
      volume         = "72",
      year           = "2012",
      pages          = "1896",
      doi            = "10.1140/epjc/s10052-012-1896-2",
      eprint         = "1111.6097",
      archivePrefix  = "arXiv",
      primaryClass   = "hep-ph",
      reportNumber   = "CERN-PH-TH-2011-297",
      SLACcitation   = "%%CITATION = ARXIV:1111.6097;%%"
}

@article{Ball:2014uwa,
      author         = "Ball, Richard D. and others",
      title          = "{Parton distributions for the LHC run II}",
      collaboration  = "NNPDF",
      journal        = "JHEP",
      volume         = "04",
      year           = "2015",
      pages          = "040",
      doi            = "10.1007/JHEP04(2015)040",
      eprint         = "1410.8849",
      archivePrefix  = "arXiv",
      primaryClass   = "hep-ph",
      reportNumber   = "EDINBURGH-2014-15, IFUM-1034-FT, CERN-PH-TH-2013-253,
                        OUTP-14-11P, CAVENDISH-HEP-14-11",
      SLACcitation   = "%%CITATION = ARXIV:1410.8849;%%"
}

@article{Sjostrand:2014zea,
      author         = "Sj{\"o}strand, Torbj{\"o}rn and Ask, Stefan and Christiansen,
                        Jesper R. and Corke, Richard and Desai, Nishita and Ilten,
                        Philip and Mrenna, Stephen and Prestel, Stefan and
                        Rasmussen, Christine O. and Skands, Peter Z.",
      title          = "{An introduction to PYTHIA 8.2}",
      journal        = "Comput. Phys. Commun.",
      volume         = "191",
      year           = "2015",
      pages          = "159",
      doi            = "10.1016/j.cpc.2015.01.024",
      eprint         = "1410.3012",
      archivePrefix  = "arXiv",
      primaryClass   = "hep-ph",
      reportNumber   = "LU-TP-14-36, MCNET-14-22, CERN-PH-TH-2014-190,
                        FERMILAB-PUB-14-316-CD, DESY-14-178, SLAC-PUB-16122,
                        --FERMILAB-PUB-14-316-CD",
      SLACcitation   = "%%CITATION = ARXIV:1410.3012;%%"
}

@Article{Alioli:2010xd,
      author         = "Alioli, Simone and Nason, Paolo and Oleari, Carlo and Re,
                        Emanuele",
      title          = "{A general framework for implementing NLO calculations in
                        shower Monte Carlo programs: the POWHEG BOX}",
      journal        = "JHEP",
      volume         = "06",
      pages          = "043",
      doi            = "10.1007/JHEP06(2010)043",
      year           = "2010",
      eprint         = "1002.2581",
      archivePrefix  = "arXiv",
      primaryClass   = "hep-ph",
}

@Article{Gleisberg:2008fv,
      author         = "Gleisberg, Tanju and H{\"o}che, Stefan",
      title          = "{Comix, a new matrix element generator}",
      journal        = "JHEP",
      volume         = "12",
      pages          = "039",
      doi            = "10.1088/1126-6708/2008/12/039",
      year           = "2008",
      eprint         = "0808.3674",
      archivePrefix  = "arXiv",
      primaryClass   = "hep-ph",
}

@article{Bothmann:2019yzt,
      author = "Bothmann, Enrico and others",
      title = "{Event generation with Sherpa 2.2}",
      journal = "SciPost Phys.",
      volume = "7",
      year = "2019",
      number = "3",
      pages = "034",
      doi = "10.21468/SciPostPhys.7.3.034",
      reportNumber = "FERMILAB-PUB-19-218-T, SLAC-PUB-17433, IPPP/19/42, MCNET-19-11",
      eprint = "1905.09127",
      archivePrefix = "arXiv",
      primaryClass = "hep-ph",
}

@article{Cascioli:2011va,
      author         = "Cascioli, Fabio and Maierh{\"o}fer, Philipp and Pozzorini, Stefano",
      title          = "{Scattering Amplitudes with Open Loops}",
      journal        = "Phys. Rev. Lett.",
      volume         = "108",
      year           = "2012",
      pages          = "111601",
      doi            = "10.1103/PhysRevLett.108.111601",
      eprint         = "1111.5206",
      archivePrefix  = "arXiv",
      primaryClass   = "hep-ph",
      reportNumber   = "ZU-TH-23-11, LPN11-66",
      SLACcitation   = "%%CITATION = ARXIV:1111.5206;%%"
}

@article{Buccioni:2019sur,
    author = "Buccioni, Federico and Lang, Jean-Nicolas and Lindert, Jonas M. and Maierh{\"o}fer, Philipp and Pozzorini, Stefano and Zhang, Hantian and Zoller, Max F.",
    title = "{OpenLoops 2}",
    eprint = "1907.13071",
    archivePrefix = "arXiv",
    primaryClass = "hep-ph",
    reportNumber = "IPPP/19/62, FR-PHENO-2019-12, PSI-PR-19-15, ZU-TH 37/19",
    doi = "10.1140/epjc/s10052-019-7306-2",
    journal = "Eur. Phys. J. C",
    volume = "79",
    number = "10",
    pages = "866",
    year = "2019"
}

@article{Denner:2016kdg,
      author         = "Denner, Ansgar and Dittmaier, Stefan and Hofer, Lars",
      title          = "{\textsc{Collier}: A fortran-based complex one-loop library in
                        extended regularizations}",
      journal        = "Comput. Phys. Commun.",
      volume         = "212",
      year           = "2017",
      pages          = "220-238",
      doi            = "10.1016/j.cpc.2016.10.013",
      eprint         = "1604.06792",
      archivePrefix  = "arXiv",
      primaryClass   = "hep-ph",
      reportNumber   = "FR-PHENO-2016-003, ICCUB-16-016",
      SLACcitation   = "%%CITATION = ARXIV:1604.06792;%%"
}

@article{Agostinelli:2002hh,
      author         = "{GEANT4 Collaboration} and Agostinelli, S. and others",
      title          = "{\textsc{Geant4} -- a simulation toolkit}",
      journal        = "Nucl. Instrum. Meth. A",
      volume         = "506",
      year           = "2003",
      pages          = "250",
      doi            = "10.1016/S0168-9002(03)01368-8",
      reportNumber   = "SLAC-PUB-9350, FERMILAB-PUB-03-339",
      SLACcitation   = "%%CITATION = NUIMA,A506,250;%%"
}


%% file: ATLAS.bib
@Article{PERF-2007-01,
    author         = "{ATLAS Collaboration}",
    title          = "{The ATLAS Experiment at the CERN Large Hadron Collider}",
    journal        = "JINST",
    volume         = "3",
    year           = "2008",
    pages          = "S08003",
    doi            = "10.1088/1748-0221/3/08/S08003",
    primaryClass   = "hep-ex",
}

@Article{SOFT-2010-01,
    author         = "{ATLAS Collaboration}",
    title          = "{The ATLAS Simulation Infrastructure}",
    journal        = "Eur. Phys. J. C",
    volume         = "70",
    year           = "2010",
    pages          = "823",
    doi            = "10.1140/epjc/s10052-010-1429-9",
    eprint         = "1005.4568",
    archivePrefix  = "arXiv",
    primaryClass   = "physics.ins-det",
}

@Article{PERF-2015-10,
    author         = "{ATLAS Collaboration}",
    title          = "{Muon reconstruction performance of the ATLAS detector in proton--proton collision data at \(\sqrt{s} = 13\,\text{TeV}\)}",
    journal        = "Eur. Phys. J. C",
    volume         = "76",
    year           = "2016",
    pages          = "292",
    doi            = "10.1140/epjc/s10052-016-4120-y",
    reportNumber   = "CERN-EP-2016-033",
    eprint         = "1603.05598",
    archivePrefix  = "arXiv",
    primaryClass   = "hep-ex",
}

@Article{STDM-2015-03,
    author         = "{ATLAS Collaboration}",
    title          = "{Measurement of \(W^{\pm}\) and \(Z\)-boson production cross sections in \(pp\) collisions at \(\sqrt{s} = 13\,\text{TeV}\) with the ATLAS detector}",
    journal        = "Phys. Lett. B",
    volume         = "759",
    year           = "2016",
    pages          = "601",
    doi            = "10.1016/j.physletb.2016.06.023",
    reportNumber   = "CERN-EP-2016-069",
    eprint         = "1603.09222",
    archivePrefix  = "arXiv",
    primaryClass   = "hep-ex",
}

@Article{PERF-2016-07,
    author         = "{ATLAS Collaboration}",
    title          = "{Performance of missing transverse momentum reconstruction with the ATLAS detector using proton--proton collisions at \(\sqrt{s} = 13\,\text{TeV}\)}",
    journal        = "Eur. Phys. J. C",
    volume         = "78",
    year           = "2018",
    pages          = "903",
    doi            = "10.1140/epjc/s10052-018-6288-9",
    reportNumber   = "CERN-EP-2017-274",
    eprint         = "1802.08168",
    archivePrefix  = "arXiv",
    primaryClass   = "hep-ex",
}

@Article{TRIG-2016-01,
    author         = "{ATLAS Collaboration}",
    title          = "{Performance of the ATLAS trigger system in 2015}",
    journal        = "Eur. Phys. J. C",
    volume         = "77",
    year           = "2017",
    pages          = "317",
    doi            = "10.1140/epjc/s10052-017-4852-3",
    reportNumber   = "CERN-EP-2016-241",
    eprint         = "1611.09661",
    archivePrefix  = "arXiv",
    primaryClass   = "hep-ex",
}

@Article{EGAM-2018-01,
    author         = "{ATLAS Collaboration}",
    title          = "{Electron and photon performance measurements with the ATLAS detector using the 2015--2017 LHC proton--proton collision data}",
    journal        = "JINST",
    volume         = "14",
    year           = "2019",
    pages          = "P12006",
    doi            = "10.1088/1748-0221/14/12/P12006",
    reportNumber   = "CERN-EP-2019-145",
    eprint         = "1908.00005",
    archivePrefix  = "arXiv",
    primaryClass   = "hep-ex",
}

@Article{FTAG-2018-01,
    author         = "{ATLAS Collaboration}",
    title          = "{ATLAS \(b\)-jet identification performance and efficiency measurement with \(t\bar{t}\) events in \(pp\) collisions at \(\sqrt{s} = 13\,\text{TeV}\)}",
    journal        = "Eur. Phys. J. C",
    volume         = "79",
    year           = "2019",
    pages          = "970",
    doi            = "10.1140/epjc/s10052-019-7450-8",
    reportNumber   = "CERN-EP-2019-132",
    eprint         = "1907.05120",
    archivePrefix  = "arXiv",
    primaryClass   = "hep-ex",
}

@Article{PIX-2018-001,
    author         = "Abbott, B. and others",
    title          = "{Production and integration of the ATLAS Insertable B-Layer}",
    journal        = "JINST",
    volume         = "13",
    year           = "2018",
    pages          = "T05008",
    doi            = "10.1088/1748-0221/13/05/T05008",
    eprint         = "1803.00844",
    archivePrefix  = "arXiv",
    primaryClass   = "physics.ins-det",
}

@Article{TRIG-2018-01,
    author         = "{ATLAS Collaboration}",
    title          = "{Performance of the ATLAS muon triggers in Run~2}",
    journal        = "JINST",
    volume         = "15",
    year           = "2020",
    pages          = "P09015",
    doi            = "10.1088/1748-0221/15/09/p09015",
    reportNumber   = "CERN-EP-2020-031",
    eprint         = "2004.13447",
    archivePrefix  = "arXiv",
    primaryClass   = "hep-ex",
}

@Article{TRIG-2018-05,
    author         = "{ATLAS Collaboration}",
    title          = "{Performance of electron and photon triggers in ATLAS during LHC Run~2}",
    journal        = "Eur. Phys. J. C",
    volume         = "80",
    year           = "2020",
    pages          = "47",
    doi            = "10.1140/epjc/s10052-019-7500-2",
    reportNumber   = "CERN-EP-2019-169",
    eprint         = "1909.00761",
    archivePrefix  = "arXiv",
    primaryClass   = "hep-ex",
}

@Article{EXOT-2020-28,
    author         = "{ATLAS Collaboration}",
    title          = "{Search for lepton-flavor-violation in \(Z\)-boson decays with \(\tau\)-leptons with the ATLAS detector}",
    journal        = "Phys. Rev. Lett.",
    volume         = "127",
    year           = "2021",
    pages          = "271801",
    doi            = "10.1103/PhysRevLett.127.271801",
    reportNumber   = "CERN-EP-2021-067",
    eprint         = "2105.12491",
    archivePrefix  = "arXiv",
    primaryClass   = "hep-ex",
}

@Booklet{ATL-SOFT-PUB-2021-003,
    author         = "{ATLAS Collaboration}",
    title          = "{ATLAS Computing Acknowledgements}",
    howpublished   = "{ATL-SOFT-PUB-2021-003}",
    url            = "https://cds.cern.ch/record/2776662",
}

@Report{ATLAS-TDR-2010-19,
    author         = "{ATLAS Collaboration}",
    title          = "{ATLAS Insertable B-Layer Technical Design Report}",
    type           = "ATLAS-TDR-19; CERN-LHCC-2010-013",
    year           = "2010",
    url            = "https://cds.cern.ch/record/1291633",
    related        = "ATLAS-TDR-2010-19-addm",
    relatedstring  = "Addendum:",
}


%% file: ConfNotes.bib
@Booklet{ATLAS-CONF-2017-029,
    author         = "{ATLAS Collaboration}",
    title          = "{Measurement of the tau lepton reconstruction and identification performance in the ATLAS experiment using \(pp\) collisions at \(\sqrt{s} = 13~\text{TeV}\)}",
    howpublished   = "{ATLAS-CONF-2017-029}",
    url            = "https://cds.cern.ch/record/2261772",
    year           = "2017",
}

@Booklet{ATLAS-CONF-2018-023,
    author         = "{ATLAS Collaboration}",
    title          = "{\(E_{\text{T}}^{\text{miss}}\) performance in the ATLAS detector using 2015--2016 LHC \(pp\) collisions}",
    howpublished   = "{ATLAS-CONF-2018-023}",
    url            = "https://cds.cern.ch/record/2625233",
    year           = "2018",
}

@Booklet{ATLAS-CONF-2019-021,
    author         = "{ATLAS Collaboration}",
    title          = "{Luminosity determination in \(pp\) collisions at \(\sqrt{s} = 13\,\text{TeV}\) using the ATLAS detector at the LHC}",
    howpublished   = "{ATLAS-CONF-2019-021}",
    url            = "https://cds.cern.ch/record/2677054",
    year           = "2019",
}


%% file: PubNotes.bib
@Booklet{ATL-PHYS-PUB-2019-033,
    author         = "{ATLAS Collaboration}",
    title          = "{Identification of hadronic tau lepton decays using neural networks in the ATLAS experiment}",
    howpublished   = "{ATL-PHYS-PUB-2019-033}",
    url            = "https://cds.cern.ch/record/2688062",
    year           = "2019",
}
